# Quantum Superposition States of Two Valleys in Graphene

Jia-Bin Qiao, Zhao-Dong Chu, Liang-Mei Wu, Lin He*

Department of Physics, Beijing Normal University, Beijing, 100875, People's Republic of China

**A system in a quantum superposition of distinct states usually exhibits many peculiar behaviors. Here we show that putting quasiparticles of graphene into superpositions of states in the two valleys, *K* and *K′*, can complete change the properties of the massless Dirac fermions. Due to the coexistence of both the quantum and relativistic characteristics, the superposition states exhibit many oddball behaviors in their chiral tunneling process. We further demonstrate that a recently observed line defect in graphene could be used to generate such superposition states. A possible experimental device to detect the novel behaviors of the relativistic superposition states in graphene is proposed.**

The superposition principle lies at the heart of quantum mechanics. Since Schrödinger's thought experiment about the unfortunate cat [1], many efforts have been spent on the realization of a quantum superposition of well separated quasi-classical states [2-7]. Such superposition states are very important not only in fundamental tests of quantum theory, but also in quantum information science, in which they are used as a central resource [8]. Several years ago, Bermudez, *et al* proposed the generation of superposition states in relativistic Landau levels when a perpendicular magnetic field couples to a relativistic spin 1/2 charged particle [9]. Their result provides us the first relativistic extension of the nonrelativistic "Schrödinger's cat". It is well known that graphene is studied widely in connection to the Dirac equation [10-16] and the Klein paradox–one of the most exotic result of quantum electrodynamics–was tested experimentally in it [17,18]. Therefore, it could be possible to realize a relativistic superposition of distinct states in this condensed-matter system.

In this Letter, we show that this is indeed possible even without applying an external magnetic field. Our analysis points out that the massless Dirac fermions of graphene can be put into a superposition of two valley states (one in the $K$ valley, the other in the $K'$ valley) by taking advantage of a recently observed line defect [19-21]. The coexistence of both the quantum and relativistic characteristics of the superposition states results in many counterintuitive properties of the system.

Many of graphene's unique electronic properties are a consequence of graphene's two-dimensional (2D) honeycomb lattice. Its crystal structure results in two

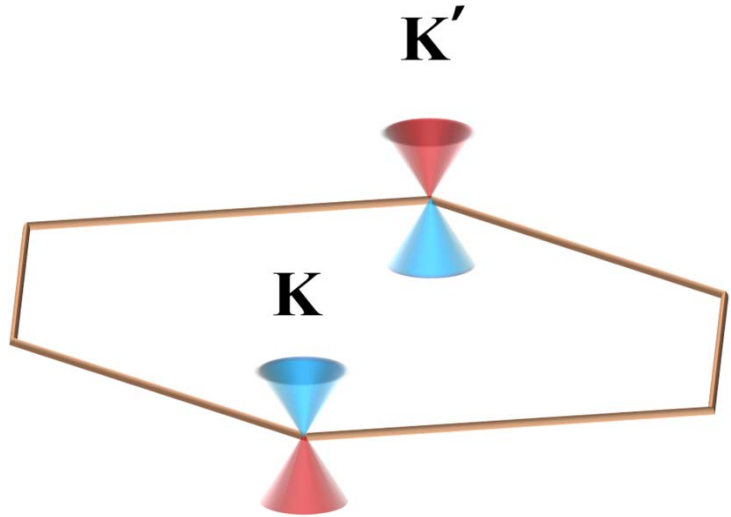

**FIG. 1 (color online).** Schematic band structure of graphene. Electronic spectrums in the vicinity of two Dirac points, *K* and *K'*, at the diagonal corners of the first Brillouin Zone (the hexagon). Two Dirac cones of the zone corners are inequivalent.

independent Dirac cones, commonly called *K* and *K'*, centered at the opposite corners of the hexagonal Brillouin zone, as shown in Fig. 1. The Dirac spinor of the two cones in graphene has the form as [12, 13]:

$$\left| K_\tau \right\rangle = \frac{1}{\sqrt{2}} \begin{pmatrix} 1 \\ \pm i e^{-i\tau\theta_\tau} \end{pmatrix}. \qquad (1)$$

Here $\tau = \pm 1$ is the valley index (+1 for the *K* valley and -1 for the *K'* valley), $v_F$ is the Fermi velocity, $\sigma_{x,y}$ is the Pauli matrix, $\theta_\tau = \arctan\left(\frac{q_{\tau,y}}{q_{\tau,x}}\right)$ is defined as the angle of wave vector $\bm{q}_\tau \equiv (q_{\tau,x}, q_{\tau,y})$ in momentum space, and the plus (minus) sign describes the electron (hole). Obviously, the two valleys are two disjoint low-energy regions in the reciprocal primitive cell and the valley index of low-energy quasiparticles in graphene usually is either +1 or -1, as shown in Fig. 1. Therefore, quasiparticles in the two Dirac cones are suggested as carriers of information in the valleytronics [22-25].

For the superposition states considered in this paper, the wave functions can be expressed as:

$$\left| K + K' \right\rangle = N\left( \left| K \right\rangle + \omega e^{i\lambda} \left| K' \right\rangle \right). \qquad (2)$$

Here *N* is a normalized constant, $\omega$ and $\lambda$ are the relative amplitude and phase difference between the state in $\left| K' \right\rangle$ and the state in $\left| K \right\rangle$, respectively. The quasiparticle in the superposition states is expected to behave distinct from its components, *i.e.*, the states in either the *K* valley or the *K'* valley, because of the quantum interference between them. Below, we will calculate the chiral or Klein tunneling to illustrate the difference between the quasiparticles described by Eq. (1)

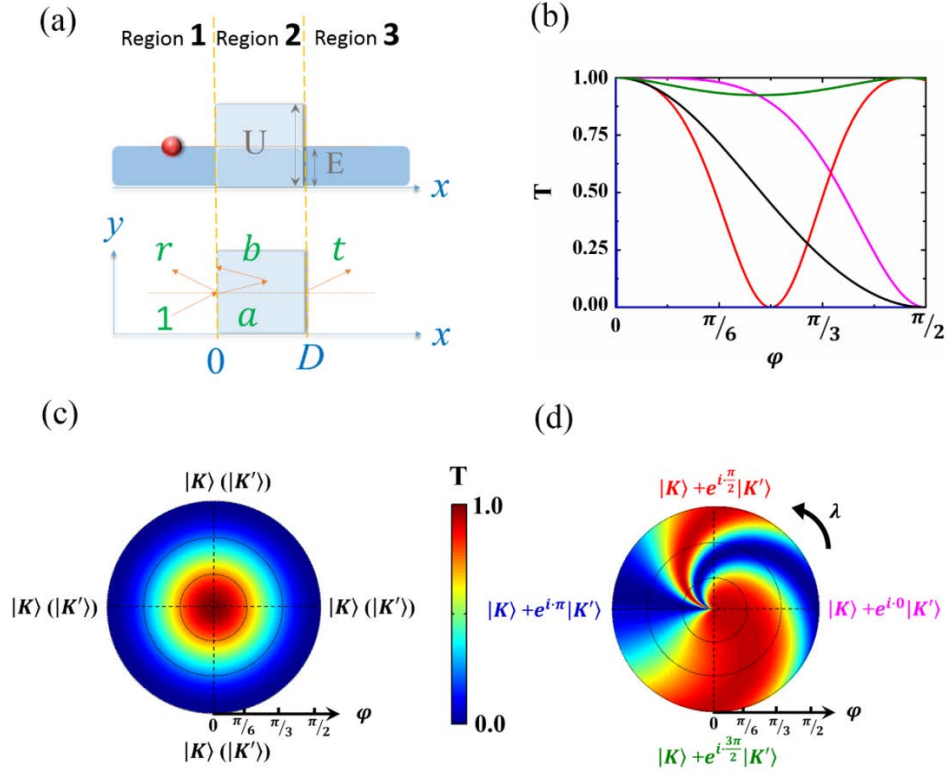

**FIG. 2 (color online).** (a) Top: Schematic diagram of an electron of energy $E$ coming to a $y$-axis-infinite barrier of height $U$ and width $D$. Bottom: Definition of some coefficients of the wave function in the three regions divided by the barrier. (b) Several examples of transmission probability $T$ through a 100-nm-wide barrier of height 450 meV as a function of the incident angle $\varphi$ for different quasiparticles in graphene. Black curve denotes the quasiparticles in a single valley (either $K$ or $K'$). For the superposition states $|K\rangle + \omega e^{i\lambda}|K'\rangle$ (here $\omega = 1$), pink curve denotes $\lambda = 0$, red curve denotes $\lambda = \pi/2$, blue curve denotes $\lambda = \pi$ and green curve denotes $\lambda = 3\pi/2$. (c) and (d): Contour map for transmission probability $T(\varphi)$ with different $\lambda$ for electron occupying the states $|K\rangle + \omega e^{i\lambda}|K'\rangle$ in graphene. In panel (c), $\omega = 0$, *i.e.*, the quasiparticles are in $K$ valley. In panel (d), $\omega = 1$.

and that described by Eq. (2).

The chiral tunneling problem of quasiparticles has been considered in several different graphene systems [12,25,26]. Figure 2(a) shows a general scheme of the tunneling system. The $y$-axis-infinite potential barrier characterized by a rectangular shape with width $D$ and height $U$ divides the system into three regions: the left of the barrier (region 1, $x < 0$), inside the barrier (region 2, $0 < x < D$), and the right of the barrier (region 3, $x > D$). It is straightforward to solve the tunneling problem with the wave functions in the three regions. For the superposition states described by Eq. (2), the wave-functions in the three regions can be written as:

$$\begin{aligned}
|\psi_1\rangle &= \left(|K(\varphi)\rangle e^{ik_x \cdot x + ik_y \cdot y} + \omega e^{i\lambda}|K'(\varphi)\rangle e^{ik_x \cdot x - ik_y \cdot y}\right) \\
&+ r\left(|K(\pi-\varphi)\rangle e^{-ik_x \cdot x + ik_y \cdot y} + \omega e^{-i\lambda}|K'(\pi-\varphi)\rangle e^{-ik_x \cdot x - ik_y \cdot y}\right), \quad x < 0 \\
|\psi_2\rangle &= a\left(|K(\theta)\rangle e^{ik_x \cdot x + ik_y \cdot y} + \omega e^{i\lambda}|K'(\theta)\rangle e^{ik_x \cdot x - ik_y \cdot y}\right) \\
&+ b\left(|K(\pi-\theta)\rangle e^{-ik_x \cdot x + ik_y \cdot y} + \omega e^{-i\lambda}|K'(\pi-\theta)\rangle e^{-ik_x \cdot x - ik_y \cdot y}\right), \quad 0 < x < D \\
|\psi_3\rangle &= t\left(|K(\varphi)\rangle e^{ik_x \cdot x + ik_y \cdot y} + \omega e^{i\lambda}|K'(\varphi)\rangle e^{ik_x \cdot x - ik_y \cdot y}\right), \quad x > D \, . \quad (3)
\end{aligned}$$

Here, $\varphi$ and $\theta$ are the incident angle and refraction angle of the quasiparticles with respect to the $x$ axis, respectively. The coefficients $r, a, b, t$, in Eq. (3) can be obtained according to the continuity of the wave functions. For the case that $\omega = 0$, $i.e.$, the quasiparticles are in the $K$ valley, then the tunneling problem becomes identical as that first considered in Ref. [12]. The transmission coefficient for the quasiparticles in the $K$ valley has the following expression (for quasiparticles in the $K'$ valley, the result is similar. See supporting materials [27] for details of calculation)

$$t_K = t_{K'} = \frac{4ss'}{\det\langle K_\tau \rangle} \cdot \cos\theta \cdot \cos\varphi, \quad (4)$$

where $s = \text{sgn}(E)$, $s' = \text{sgn}(E-U)$, and the Matrix $\langle K_\tau \rangle$ reads

$$\langle K_\tau \rangle = \begin{pmatrix} 1 & 1 & -1 & 0 \\ e^{iq_x \cdot D} & e^{-iq_x \cdot D} & 0 & -e^{ik_x \cdot D} \\ s'ie^{-i\tau\cdot\theta} & -s'ie^{i\tau\cdot\theta} & sie^{i\tau\cdot\varphi} & 0 \\ s'ie^{-i\tau\cdot\theta}e^{iq_x \cdot D} & -s'ie^{i\tau\cdot\theta}e^{-iq_x \cdot D} & 0 & -sie^{-i\tau\cdot\varphi}e^{ik_x \cdot D} \end{pmatrix}. \quad (5)$$

Figure 2(b) shows an example of the angular dependence of the transmission probability $T = |t_K|^2$ calculated using the Eq. (4) and Eq. (5). The perfect transmission for electrons incident in the normal direction of the potential barrier, *i.e.*, $\varphi = 0$, is the feature unique to massless Dirac fermions and is viewed as an incarnation of the Klein paradox [12]. Such a result, *i.e.*, the perfect transmission at $\varphi = 0$, can also be treated as a direct manifestation of the relativistic characteristic of the quasiparticles in graphene.

For the case that $\omega \neq 0$, *i.e.*, the quasiparticles are in the superposition states of the two valleys, then the transmission coefficient $t_{K+K'}$ has the following form:

$$t_{K+K'} = t_K + t_{K'} + \tilde{t} \,. \quad (6)$$

Where $\tilde{t} = \dfrac{ss'}{\det\langle m \rangle \det\langle M \rangle} \left( M_t \cdot \det\langle m \rangle - 8\cos\theta \cdot \cos\varphi \cdot \det\langle M \rangle \right)$ can be viewed as the coefficient induced by the quantum interference of the two valleys. The parameter $M_t$ is

$$\begin{aligned} M_t =\ & 2ss'\cos(\theta-\varphi)(1+6\omega^2+\omega^4+4\omega\cos\lambda+4\omega^3\cos\lambda) \\ & + 2ss'\cos(\theta+\varphi)(1+2\omega^2+\omega^4) \\ & + 8ss'\cos(\theta+\varphi+\lambda)(\omega+\omega^3) \\ & + 8ss'\omega^2\cos(\theta+\varphi+2\lambda) \end{aligned} \quad (7)$$

and the Matrix $\langle M \rangle$ can be written as

$$\langle M \rangle = \sum_{i,j}^{4} \langle K \rangle_{i,j} + \sum_{i,j=1,4} \langle K' \rangle_{i,j} \omega e^{i\lambda} + \sum_{i,j=2,3} \langle K' \rangle_{i,j} \omega e^{-i\lambda}. \quad (8)$$

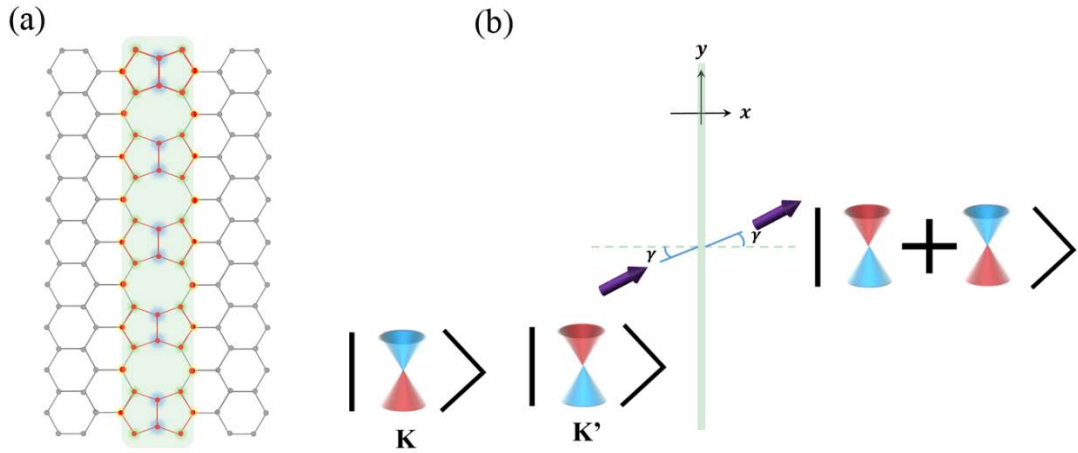

**FIG. 3 (color online).** (a) Atomic structure of graphene with a line defect, as highlighted in gray. The structure exhibits the symmetry of the two sublattices in graphene with respect to the line defect. (b) A schematic map of the generation of the superposition states using the line defect. An incoming quasiparticle approaching the line defect is in a single valley (either $K$ or $K'$). The transmitted electrons in the right of the line defect are forced into the superposition states of the two valleys.

Figure 2(b) also shows several examples of the angular dependence of the transmission probability $T = |t_{K+K'}|^2$ calculated using the Eqs. (6-8). In the calculation, we assume $\omega = 1$ for simplicity and only change the phase difference $\lambda$ between the two valleys. At $\varphi = 0$, the potential barrier is still transparent (T = 1) for all the superposition states with different $\lambda$. However, for $\varphi \neq 0$, the transmission probability of the superposition states depends sensitively on the phase difference, as shown in Fig. 2(b-d). The quantum interference of the two components during the tunneling process results in many counterintuitive behaviors of the superposition states. For example, when the phase difference $\lambda = \pi$, the transmission probability of the superposition state is 100% at $\varphi = 0$ and becomes zero at any nonzero incident angle, as shown in Fig. 2(b). It means that the destructive interference between the $K$ valley and the $K'$ valley leads to a perfect reflection of the superposition state at any nonzero incident angle. The only exception is when the incident angle is exactly zero (the nature of the quasiparticle in this superposition state is still massless Dirac fermion). As mentioned above, the perfect transmission for $\varphi = 0$ is the manifestation of the relativistic nature of the quasiparticles. Therefore, this oddball tunneling behavior, *i.e.*, the perfect transmission at $\varphi = 0$ and perfect reflection for $\varphi \neq 0$, is unique to the relativistic superposition state and should be attributed to the coexistence of the quantum and relativistic characteristics of the superposition state. Subsequently, we will demonstrate that such a superposition state is not science fiction and this counterintuitive tunneling behavior could be tested experimentally in graphene.

It is well established that the electronic spectra of graphene and the wave functions of its quasiparticles depend sensitively on the lattice structure [13-16]. More importantly, the 2D nature of graphene offers unique advantages in tailoring the arrangement of carbon atoms. Therefore, graphene provides unprecedented platform to tune the wave functions of its quasiparticles. Figure 3(a) shows an atomic structure of graphene with a line defect [19], which is predicted to be used as a valley filter in the valleytronics [20]. Very recently, it was demonstrated that such a line defect can be grown in a controlled way in graphene [21]. Below we will show that the transmitted electrons across the line defect are forced into the superposition states of the two valleys, as shown in Fig. 3(b).

As demonstrated in Ref. [20], the symmetry of the line defect plays a vital role in determining the electronic properties of the system. Around the Dirac point, the reflection operator about the line defect commutes with the Hamiltonian describing the system. Therefore, the eigenstates of the Hamiltonian approaching the line defect are either symmetric or antisymmetric with respect to reflection. There is a node at the line defect for the antisymmetric states, whereas the symmetric states don't have a node at the line defect. As a consequence, only the symmetric states can carry electrons across the line defect and contribute to the transmission. The antisymmetric states only contribute to the reflection of the line defect. The eigenstates of the transmission (symmetric) and reflection (antisymmetric) components are [20]

$$|\rightarrow\rangle = \frac{1}{\sqrt{2}}\begin{pmatrix}1\\1\end{pmatrix}, \qquad |\leftarrow\rangle = \frac{1}{\sqrt{2}}\begin{pmatrix}1\\-1\end{pmatrix}. \qquad (9)$$

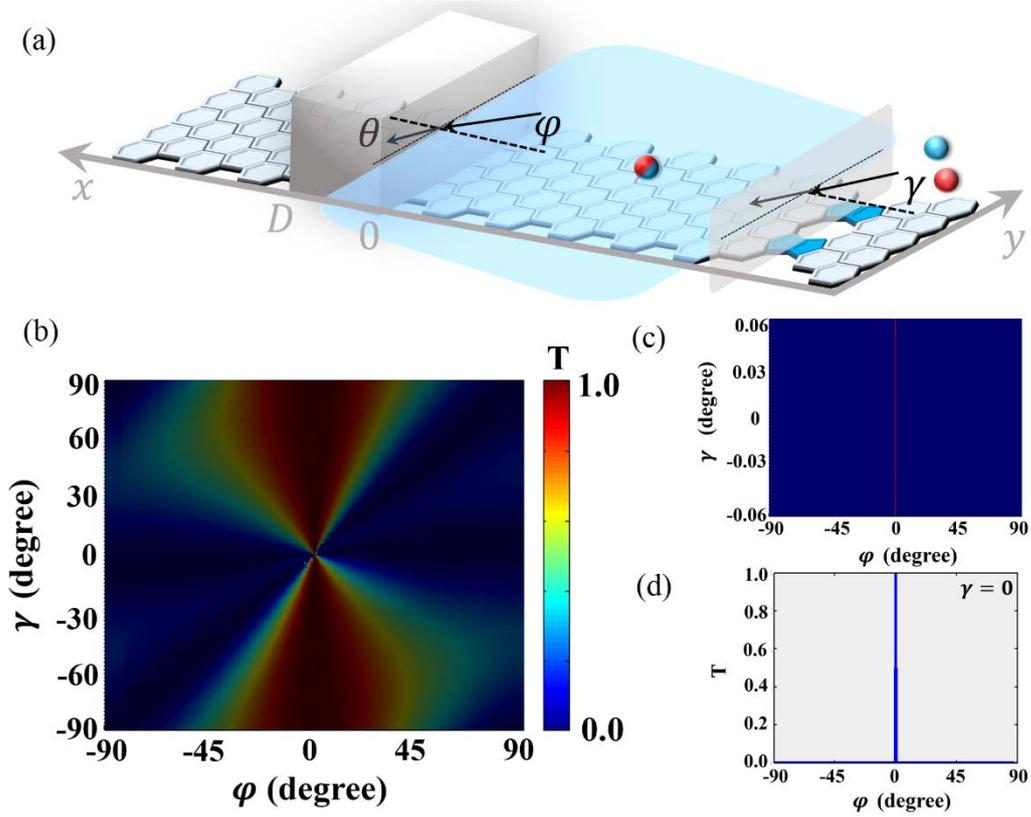

**FIG. 4 (color online).** (a) A schematic experimental structure to measure the quantum tunneling of the superposition states of the two valleys. The red (blue) ball denotes the electrons occupying the $K$ ($K'$) valley. The red-and-blue ball denotes the superposition state of electrons across the line defect. $\varphi$ and $\gamma$ are the incident angles of the potential barrier and the line defect, respectively. (b) Transmission probability as a function of the incident angle $\gamma$ and $\varphi$. (c) Zoom-in of the transmission probability of panel (b) around $\gamma = 0$. (d) Transmission probability as a function of $\varphi$ at $\gamma = 0$.

Then, the eigenstates of graphene, as described by Eq. (1), approaching the line defect can be expressed as

$$|K_\tau\rangle = \frac{1 + ie^{-i\tau\theta_\tau}}{2}|\rightarrow\rangle + \frac{1 - ie^{-i\tau\theta_\tau}}{2}|\leftarrow\rangle . \quad (10)$$

For incident electron with angle of incidence $\gamma$, as shown in Fig. 3(b), one immediately obtains the transmission components as superposition states of the two valleys

$$|\rightarrow\rangle = N_0\left(|K\rangle + \omega_0 e^{i\lambda_0}|K'\rangle\right) , \quad (11)$$

where $N_0 = \dfrac{1 - ie^{i\gamma}}{2\sin\gamma + (\sqrt{2} - \sqrt{2}ie^{i\gamma} - 2\sin\gamma)\delta_{\gamma,0}}$ is the normalized constant and $\omega_0 e^{i\lambda_0} = -\dfrac{e^{-i\cdot 2\gamma} + 1}{2 + 2\sin\gamma}$. The result of Eq. (11) indicates that the line defect can be used to generate the superposition states of the two valleys and it is facile to tune the wave functions, *i.e.*, the relative amplitude and phase difference between the two components, by adjusting the incident angles of electrons. A combination of electron supercollimation effect–an effect forces the electrons to move undistorted along a selected direction–of massless Dirac fermions [28,29] and recently developed methods in generating one-dimensional electronic superlattice in graphene [30-32] could help to control the incident angles of electrons precisely. Therefore, realization of the superposition states depicted by Eq. (11) in a controllable way is within the grasp of today's technology.

Figure 4(a) shows a schematic experimental device to measure the counterintuitive chiral tunneling of the relativistic superposition states in graphene. Two important parameters, the incident angle of the line defect $\gamma$ and the incident angle of the potential barrier $\varphi$, dominate the behaviors of the chiral tunneling, as shown in Fig. 4(b) (See supporting materials [27] for details of calculation). At a high angle of incidence to the line defect, *i.e.*, $\gamma \to +90°$, only the electrons occupying the *K* valley transmitted across the line defect (when $\gamma \to -90°$, only the electrons occupying the *K'* valley across the line defect), which is the reason why the line defect is predicted as a valley filter [20]. For this case, the chiral tunneling is almost identical to that of electrons occupying a single valley, as considered in Ref. [12] and shown in Fig. 2(c). The effect of quantum interference between the two valleys appears when the incident angle deviates from $\pm 90°$ and it becomes remarkable at small angles of incidence, as shown in Fig. 4(b). For the case that $\gamma = 0$, the wave function of the transmitted electrons across the line defect can be simplified as $\frac{\sqrt{2}}{2}(|K\rangle + e^{i\pi}|K'\rangle)$. Such a unique superposition state could exhibit oddball behaviors of quantum tunneling because of the phase difference $\pi$ of the two components. As discussed above, the transmission probability is 100% when the incident angle $\varphi$ exactly equals to zero and the destructive interference between the two valleys will result in a perfect reflection of the superposition state at any nonzero incident angle. This counterintuitive behavior is further confirmed, as shown in Fig. 4(c) and Fig. 4(d). If all the charge carriers of this system are put into such a superposition state, then the transport properties of the system can be tuned from insulating to metallic, or vice versa, by slightly changing the relative angle between the line defect and the potential barrier. This result indicates that it is possible to tune the properties of graphene using the peculiar behaviors of the superposition states.

In summary, we show that it is possible to realize relativistic superposition states in graphene–a condensed-matter system–using a line defect. The interference between the two components, the *K* and *K′* valleys, of the superposition states leads to many oddball behaviors. The counterintuitive chiral tunneling of the superposition states obtained in this paper is attributed to the coexistence of both the quantum and relativistic characteristics of the quasiparticles. An experimental device to detect the relativistic superposition states is also proposed and the predicted effect is expected to be realized in the near future.

**Acknowledgments**

We are grateful to National Key Basic Research Program of China (Grant No. 2014CB920903, No. 2013CBA01603), National Science Foundation (Grant No. 11422430, No. 11374035, No. 11004010), the program for New Century Excellent Talents in University of the Ministry of Education of China (Grant No. NCET-13-0054), and Beijing Higher Education Young Elite Teacher Project (Grant No. YETP0238).

*E-mail: helin@bnu.edu.cn